\documentclass[12pt]{article}
\usepackage{amsfonts}
\usepackage{graphicx}
\usepackage{epstopdf}
\usepackage{epsfig}
\usepackage{amssymb}
\usepackage{setspace}
\usepackage{caption}
\usepackage{amsfonts}
\usepackage{color}
\usepackage{amsmath}
\usepackage{float}
\usepackage{comment}
\newcommand {\be}{\begin{equation}}
\newcommand {\ee}{\end{equation}}
 \newcommand {\bea}{\begin{array}}
 
 \newcommand {\eea}{\end{array}}

 \newcommand {\RN}{Reissner-Nordstrom~}
 \newcommand {\sch}{Schwarzschild~}
  
\textwidth=6.5in \hoffset=-.75in \voffset=-1in \numberwithin{equation}{section}
\numberwithin{figure}{section}

\begin{document}

\begin{titlepage}
	\vspace{1cm}
	\begin{center}
		{\Large \bf {Magnetized Reissner-Nordstrom-Taub-NUT spacetime and microscopic entropy}}\\
	\end{center}
	\vspace{2cm}
	\begin{center}
		\renewcommand{\thefootnote}{\fnsymbol{footnote}}
		Haryanto M. Siahaan{\footnote{haryanto.siahaan@unpar.ac.id}}\\
		Center for Theoretical Physics,\\
		Department of Physics, Parahyangan Catholic University,\\
		Jalan Ciumbuleuit 94, Bandung 40141, Indonesia
		\renewcommand{\thefootnote}{\arabic{footnote}}
	\end{center}
	
	\begin{abstract}
		
	We present a novel solution describing magnetized spacetime outside a massive object with electric charge equipped with NUT parameter. To get the solution, we employ the Ernst magnetization to the Reissner-Nordstrom-Taub-NUT spacetime as the seed. After discussing some physical aspects of the spacetime, we show that the extremal entropy of a magnetized Reissner-Nordstrom-Taub-NUT black hole can be reproduced by using the Cardy formula. 
		
	\end{abstract}
\end{titlepage}\onecolumn
\bigskip

\section{Introduction}
\label{sec:intro}

\RN black hole is an exact solution of the Einstein-Maxwell theory describing the spacetime outside a collapsing matter with electric charge. Despite it is very unlikely for a collapsing matter to maintain a significant amount of electric charge in real world, still the \RN solution has been one of the most discussed topics in gravitational researches \cite{Griffiths:2009dfa} especially black hole related \cite{Feng:2020tyc,Wang:2020vpn,Han:2020fsa,Mo:2018nnu}. Moreover, inspired by the Kerr/CFT correspondence, the charged black hole/CFT holography has been investigated \cite{Garousi:2009zx,Chen:2010bsa,Chen:2010yu} where some properties of extremal or near extremal \RN black holes can be reproduced by using some two dimensional conformal field theory approaches. 

In addition to the black hole with electric charge, there also exist exact solutions describing black hole immersed in external magnetic field. Solution where the magnetic field is considered as some perturbations in the spacetime was introduced by Wald \cite{Wald:1974np}, and for the case of strong magnetic field was proposed by Ernst \cite{Ernst}. In the proposal by Ernst, the magnetized spacetime is obtained by using a Harrison type of transformation applied to a seed solution in Einstein-Maxwell theory. For example, one can use the Kerr-Newman solution as the seed to obtain the magnetized Kerr-Newman spacetime \cite{Ernst}. Setting the mass, electric charge, and rotation parameters in the magnetized Kerr-Newman solution yields the Melvin magnetic universe \cite{Melvin:1963qx}. Interestingly, the presence of external magnetic field does not change the area of embedded black hole, but rather to deform the surface as studied in \cite{Wild:1980zz}. Various studies on black holes in external magnetic field can be found in literature \cite{ernst-wild,Aliev:1989wx,Brito:2014nja,Bicak:2015lxa,Kolos:2015iva,Tursunov:2014loa,Astorino:2016hls,Booth:2015nwa,Gibbons:2013yq,Gibbons:2013dna,Aliev:1989wz,Aliev:1980et,Aliev:1986wu,Aliev:1988wy,Aliev:1989sw,Galtsov:1978ag,Hiscock:1980zf}.

A black hole in Einstein-Maxwell theory can have the NUT parameter in addition to the well known hairs of black hole, namely the mass $m$, rotation $a$, and electric charge $q$ \cite{Griffiths:2009dfa}. Sometime adding this NUT parameter to a known black hole solution in theory can be considered as a sort of NUT extension to the solution. For example, adding NUT parameter to the \sch spacetime yields the Taub-NUT metric of the vacuum Einstein system. One can also add NUT parameter to the Kerr-Newman solution to yield the Kerr-Newman-Taub-NUT spacetime. The Taub-NUT solution mentioned previously is just a neutral and static limits of Kerr-Newman-Taub-NUT family. Despite the obscure realization of NUT parameter in our real world, solutions containing this quantity have contributed significantly in the gravitational studies. For example the motion of charged test particle was studied in \cite{Cebeci:2015fie}, and the gravitomagnetism in this spacetime was investigated in \cite{Bini:2003bm}.  Spacetime with NUT parameter can also exist in gravitational theories beyond Einstein-Maxwell, for example in scalar-tensor \cite{Cisterna:2021xxq}, Randall-Sundrum braneworld \cite{Siahaan:2020bga}, and low energy heterotic string \cite{Siahaan:2019kbw} theories. In a recent work \cite{Ciambelli:2020qny}, the authors show that the Misner string contribution to the Taub-NUT-AdS entropy can be renormalized by introducing the Gauss-Bonnet term.

Particularly, related to the seed solution to be magnetized in this paper, we consider the \RN-Taub-NUT (RNTN) solution of Einstein-Maxwell theory describing spacetime outside static charged collapsing object with NUT parameter. One can also consider this as a Taub-NUT extension of the \RN solution. The magnetized \RN spacetime had been reported decades ago \cite{Wild:1980zz}, where the corresponding Kerr/CFT correspondence study was carried out in \cite{Astorino:2015lca}. 

Here we would like to obtain the NUT extension of a magnetized \RN black hole. To proceed, we employ the Ernst magnetization transformation to the RNTN spacetime as the seed. A sort of similar problem was investigated in \cite{Frolov:2017bdq}, where the authors discussed a spacetime with NUT parameter surrounded by some weak magnetic fields. Aspects of the obtained spacetime solution in this paper are discussed, such as the black hole area and Hawking temperature. We then test the conjectured extremal Kerr/CFT correspondence in the new spacetime solution, where we are able to recover the extremal entropy by using the asymptotic symmetry group method \cite{Guica:2008mu,Hartman:2008pb,Compere:2012jk}. 

The organization of this paper is as follows. In the next section, we construct the magnetized RNTN solution by employing the Ernst magnetization to the RNTN spacetime as the seed. The area, entropy, and Hawking temperature of the black hole in the spacetime are discussed in section \ref{sec.aspects}. In section \ref{sec.kerrcft}, we compute the entropy of extremal MRNTN black hole using microscopic formula. Finally we give conclusions and discussions. We consider the natural units $c={\hbar} = k_B = G_4 = 1$.

\section{Construction of magnetized Reissner-Nordstrom-Taub-NUT spacetime}
\label{sec:MRNTNconstruction}

\subsection{Ernst magnetization}

Here we review the Ernst magnetization prescription, which basically is a type of Harrison transformation. We consider the following stationary and axial symmetric line element, 
\be\label{metricLPW} 
{\rm{d}}s^2  = f\left( {\omega {\rm{d}}t-{\rm{d}}\phi } \right)^2 + f^{ - 1} \left( { e^{2\gamma } {\rm{d}}\chi {\rm{d}}{\bar \chi} -\rho ^2 {\rm{d}}t^2 } \right) \,,
\ee
which is known as the Lewis-Papapetrou-Weyl (LPW) type of metric. In the metric above, the functions $f$, $\gamma$, and $\omega$ depend on $\chi$, and we have used the $-+++$ signs convention for the spacetime. Now let us also consider two kind of complex potentials, namely
\be \label{Ernst.EM}
\Phi  = A_\phi   + i\tilde A_\phi\,,
\ee  
and
\be \label{Ernst.grav}
{\cal E} = f + {{\Phi }{\bar \Phi }}   - i\Psi \,.
\ee 
These complex functions are known as the electromagnetic and gravitational Ernst potentials, respectively. Note that the real part of $\Phi$ above is $A_\phi$ is a component in the vector potential $A_\mu$, and the imaginary part comes from the vector field ${\tilde A}_\mu$ which builds the dual field strength tensor ${{\tilde F}_{\mu \nu }} = {\partial _\mu }{{\tilde A}_\nu } - {\partial _\nu }{{\tilde A}_\mu }$ where ${{\tilde F}_{\mu \nu }} =\tfrac{1}{2} {\varepsilon _{\mu \nu \alpha \beta }}{F^{\alpha \beta }}$. To find the solution for $A_t$, we can use the following equation 
\be \label{eqAtilde}
-i\frac{\rho }{f}\nabla \tilde A_\phi  = \nabla A_t  +\omega \nabla A_\phi \,.
\ee 
The twist potential $\Psi$ in gravitational potential (\ref{Ernst.grav}) is related to the metric functions and vector potential by the equation
\be \label{eq.Psi}
- i\nabla \Psi  = \frac{{ f^2 }}{\rho }\nabla \omega + 2{\bar \Phi} \nabla \Phi  \,.
\ee 

Using these two potentials above, the Einstein-Maxwell equations
\be \label{eq.Einstein-Maxwell}
{R_{\mu \nu }} = 2{F_{\mu \alpha }}F_\nu ^\alpha  - \frac{1}{2}{g_{\mu \nu }}{F_{\alpha \beta }}{F^{\alpha \beta }}\,,
\ee  
can be rewritten in the form of wave equations
\be \label{eq.Ernst.grav}
\left( {{\cal E} +{\bar {\cal E}} + {\Phi {\bar \Phi}}} \right)\nabla^2 {\cal E} = 2\left( {\nabla {\cal E} + 2{\bar \Phi}\nabla \Phi } \right) \cdot \nabla {\cal E}\,,
\ee 
\be \label{eq.Ernst.EM}
\left( {{\cal E} +{\bar{\cal E}} + {\Phi {\bar \Phi}}} \right)\nabla^2 {\Phi} = 2\left( {\nabla {\cal E} + 2{\bar{\Phi}}\nabla \Phi } \right) \cdot \nabla {\Phi}\,,
\ee 
known as the Ernst equations \cite{Griffiths:2009dfa}.

There exist some transformations that leave the two equations (\ref{eq.Ernst.grav}) and (\ref{eq.Ernst.EM}) above to be invariant. One example is the Harrison transformation \cite{Harrison}. Ernst showed that one can perform a magnetization of a known spacetime using a type of Harrison transformation, namely 
\be \label{magnetization}
{\cal E} \to {\cal E}' = \Lambda ^{ - 1} {\cal E}~~~{\rm and}~~~\Phi  \to \Phi ' = \Lambda ^{ - 1} \left( {\Phi  - b {\cal E}} \right)\,,
\ee
where 
\be \label{LambdaDEF}
\Lambda  = 1 - 2b\Phi  + b^2 {\cal E}\,.
\ee 
Above, the constant $b$ represents the strength of external magnetic field in the spacetime\footnote{Here we use a quite different label for the strength of magnetic field, where in some other works \cite{Siahaan:2015xia,Astorino:2015lca,Astorino:2015naa} is used $B=2b$ instead.}. The transformation (\ref{magnetization}) leaves equations (\ref{eq.Ernst.grav}) and (\ref{eq.Ernst.EM}) unchanged for the new potentials ${\cal E}'$ and $\Phi'$ from which a new set of solutions $\left\{A'_\mu,g'_{\mu\nu}\right\}$ to the Einstein-Maxwell equations can be extracted.  

It can be shown that the magnetization transformation (\ref{magnetization}) acting on the potentials ${\cal E}$ and $\Phi$ built from a seed solution lead to the metric functions fulfilling
\be \label{fp}
f' = {\rm Re}\left\{{\cal E'}\right\} - {\Phi '} {\bar {\Phi '}}  = \frac{f}{\left| \Lambda  \right|^2}\,,
\ee 
and
\be \label{wp}
\nabla \omega ' = \left| \Lambda  \right|^2 \nabla \omega  + \frac{\rho }{f}\left( {\Lambda \nabla {\bar {\Lambda}} - {\bar \Lambda} \nabla \Lambda  } \right)\,,
\ee 
while $\gamma$ remains unchanged. Since all the incorporating function in the metric (\ref{metricLPW}) depend on $\rho$ and $z$ only, then the operator $\nabla$ can be defined in the flat Euclidean space
\be\label{metric2rho.z}
{\rm{d}}\chi {\rm{d}}{\bar {\chi}}  = {\rm{d}}\rho ^2  + {\rm{d}}z^2 \,,
\ee 
where we have set the complex coordinate $d\chi  = d\rho  + idz$ and operator $\nabla  = \partial _\rho   + i\partial _z $ accordingly. 

However, a typical spacetime solution in Einstein-Maxwell theory containing black hole is expressed by using the Boyer-Lindquist type coordinate $\left\{ {t,r,x = \cos \theta ,\phi } \right\}$. Surely we can bring the LPW line element (\ref{metricLPW}) into a Boyer-Lindquist type, where now we have
\be \label{metric2rx}
{\rm{d}}\chi {\rm{d}}{\bar{\chi}}  = \frac{{{\rm{d}}r^2 }}{{\Delta _r }} + \frac{{{\rm{d}}x^2 }}{{\Delta _x }}\,,
\ee 
with $\Delta _r = \Delta _r \left(r\right)$ and $\Delta _x = \Delta _x \left(x\right)$. Accordingly, the corresponding operator $\nabla$ will read $\nabla  = \sqrt {\Delta _r } \partial _r  + i\sqrt {\Delta _x } \partial _x $, and since $\rho^2 = \Delta_r\Delta_x$ we then have
\be \label{drAt}
\partial _r A_t  = \frac{{\Delta _x }}{f}\partial _x \tilde A_\phi  - \omega \partial _r A_\phi  \,,
\ee 
and
\be \label{dxAt}
-\partial _x A_t  = \frac{{\Delta _r }}{f}\partial _r \tilde A_\phi + \omega \partial _x A_\phi  \,,
\ee 
from eq. (\ref{eqAtilde}). The last two equations are useful later in obtaining the $A_t$ component associated to the magnetized spacetime according to (\ref{magnetization}). To end some details on magnetization procedure, another equations that we may require to complete the magnetized solution are
\be \label{drwp}
\partial _r \omega ' = \left| \Lambda  \right|^2 \partial _r \omega  - \frac{{\Delta _x }}{f}{\mathop{\rm Im}\nolimits} \left\{ { \Lambda \partial _x {\bar{\Lambda}} -{\bar{\Lambda}} \partial _x \Lambda  } \right\} \,,
\ee 
and
\be \label{dxwp}
\partial _x \omega ' = \left| \Lambda  \right|^2 \partial _x \omega  + \frac{{\Delta _r }}{f}{\mathop{\rm Im}\nolimits} \left\{ {\Lambda \partial _r {\bar{\Lambda}} - {\bar{\Lambda}} \partial _r \Lambda } \right\}\,.
\ee 
In the following section, the Ernst magnetization procedure reviewed here will be applied to the RNTN solution, to obtain the magnetized version of the solution.

\subsection{Magnetized Reissner-Nordstrom-Taub-NUT spacetime}

We start with the metric
\be \label{metricRNTN}
{\rm{d}}{s^2} =  - {{{\Delta _r}{{\left( {{\rm{d}}t + 2lx{\rm{d}}\phi } \right)}^2}} \over {{r^2} + {l^2}}} + \left( {{r^2} + {l^2}} \right){\Delta _x}{\rm{d}}{\phi ^2} + \left( {{r^2} + {l^2}} \right)\left( {{{{\rm{d}}{r^2}} \over {{\Delta _r}^2}} + {{{\rm{d}}{x^2}} \over {{\Delta _x}^2}}} \right)\,,
\ee 
where ${\Delta _r} = {r^2} - 2mr + {q^2} - {l^2}$ and ${\Delta _x} = 1 - {x^2}$. The metric above together with the vector potential
\be \label{ARNTN}
{A_\mu }{\rm{d}}{x^\mu } = {{qr\left({\rm{d}}t+2lx {\rm{d}}\phi\right)} \over {{r^2} + {l^2}}}\,,\ee  
solve the Einstein-Maxwell equation (\ref{eq.Einstein-Maxwell}). The line element (\ref{metricRNTN}) and vector (\ref{ARNTN}) are known as the RNTN solution.

To get the magnetized version of the spacetime solution above, we need to bring the seed metric (\ref{metricRNTN}) into the LPW form (\ref{metricLPW}). In doing this, we have the associated metric functions
\be 
f  = \frac{{\left( {r^2  + l^2 } \right)^2  + x^2 \left( {3l^4  - r^4  + l^2 \left\{ {8mr - 6r^2  - 4q^2 } \right\}} \right)}}{{\left( {r^2  + l^2 } \right)}}\,,
\ee 
\be 
\omega  = \frac{{2lx\Delta _r }}{{\left( {r^2  + l^2 } \right)^2  + x^2 \left( {3l^4  - r^4  + l^2 \left\{ {8mr - 6r^2  - 4q^2 } \right\}} \right)}}\,,
\ee
\be 
\rho ^2  = \Delta _x \Delta _r \,,
\ee 
and
\be 
e^{2\gamma } f^{-1}  =  r^2  + l^2\,.
\ee
With this functions and vector solution (\ref{ARNTN}), the corresponding gravitational potential is
\be \label{Ernst.grav.seed}
{\cal E}  = \frac{{l^3 \left( {1 + 3x^2 } \right) + il^2 \left( {3r\Delta _x  + 2mx^2 } \right) + l\left( {6mrx^2  - 3x^2 \left\{ {r^2  + q^2 } \right\} - r^2 } \right) + ir\left( {q^2 x^2  + r^2 \Delta _x } \right)}}{{l + ir}}\,,
\ee 
while the electromagnetic one reads
\be \label{Ernst.EM.seed}
\Phi  = \frac{{ - iqx\left( {ir - l} \right)}}{{ir + l}}\,.
\ee 
Furthermore,  from (\ref{LambdaDEF}), these two potentials give us
\[
\Lambda  = 1 + \frac{{i2bqx\left( {ir - l} \right)}}{{ir + l}} + \frac{{b^2 }}{{ir + l}}\left\{ {l^3 \left( {1 + 3x^2 } \right) + il^2 \left( {3r\Delta _x  + 2mx^2 } \right)} \right.
\]
\be \label{Lambda}
\left. { + l\left( {6mrx^2  - 3x^2 \left\{ {r^2  + q^2 } \right\} - r^2 } \right) + ir\left( {q^2 x^2  + r^2 \Delta _x } \right)} \right\}\,.
\ee 
Finally, we can get the magnetized Ernst potentials ${\cal E}' $ and $\Phi '$ after inserting (\ref{Ernst.grav.seed}), (\ref{Ernst.EM.seed}), and (\ref{Lambda})
into (\ref{magnetization}). The resulting line element can be written as
\be\label{magLPWmetric} 
{\rm{d}}s^2  = \frac{1}{{f'}}\left\{ { - \rho ^2 {\rm{d}}t^2  + e^{2\gamma } \left( {\frac{{{\rm{d}}r^2 }}{{\Delta _r }} + \frac{{{\rm{d}}x^2 }}{{\Delta _x }}} \right)} \right\} + f'\left( {{\rm{d}}\phi  - \omega '{\rm{d}}t} \right)^2  \,,
\ee  
where the functions appearing in metric above read
\be \label{ff}
f' = \frac{\left\{ {\left( {r^2  + l^2 } \right)^2  + x^2 \left( {3l^4  - r^4  + l^2 \left\{ {8mr - 6r^2  - 4q^2 } \right\}} \right)} \right\}}{\sum\limits_{k = 0}^6 {{\bar f}_k l^k } }\,,
\ee 
where
\[ 
{\bar f}_0 = r^2 \left[ {b^4 r^4 \Delta _x^2  + 2b^2 r^2 \Delta _x \left( {1 + b^2 q^2 x^2 } \right) + 1 + b^2 q^2 x^2 \left( {6 + b^2 q^2 x^2 } \right)} \right]\,,
\] 
\[ 
{\bar f}_1 = -8bqxr \left(1 -{b}^{2}{x}^{2}{q}^{2} -2{b}^{2}{r}^{2}{x}^{2}+3{b}
^{2}mr{x}^{2} \right)\,,
\] 
\[ 
{\bar f}_2 = 1+{b}^{2}{x}^{2}{q}^{2} \left( 9{b}^{2}{x}^{2}{q}^{2}-2 \right)+ r^4 {b}^{4} \left( 15{x}^{4}+7-6{x}^{2} \right) -8r^3 {b}^{4}m{x}^{2} \left( 5{x}^{2}+1 \right) 
\]
\[ 
+4 r^2 {b}^{2} \left( 3{b}^{2}{x}^{2}{q}^{2}+1-3{x}^{2}+3{b}^{2}{x}^
{4}{q}^{2}+9{b}^{2}{m}^{2}{x}^{4} \right) -16r {b}^{2}m{x}^{2} \left( 2{b}^{2}{x}^{2}{q}^{2}-1 \right)\,,
\] 
\[ 
{\bar f}_3 = -8b^3 qx \left(mx^2+2r\right) \,,
\] 
\[ 
{\bar f}_4 = b^2 \left[4b^2m^2x^4+24b^2 r x^2 \Delta_x m -9b^2 x^4 \left(r^2 +2q^2\right) - 6x^2 \left(q2b^2+5b^2r^2-1\right) +7b^2 r^2 +2\right]\,,
\] 
\[ 
{\bar f}_5 = 0\,,
\] 
\[ 
{\bar f}_6 = b^4 \left(1+3x^2\right)^2\,,
\]
and
\be  \label{wf}
\omega ' = \frac{{\sum\limits_{k = 0}^5 {{\bar\omega}_k x^k } }}{{\left( {r^2  + l^2 } \right)^2  + x^2 \left( {3l^4  - r^4  + l^2 \left\{ {8mr - 6r^2  - 4q^2 } \right\}} \right)}}\,,
\ee
where
\[
{\bar\omega}_5 = {b}^{4}l  \left(4{l}^{2}{m}^{2}+4{r}^{3}m-4{l}^{2}mr-{r}^{4}+6{r}^{2}{l}^{2}-4{q}^{2}{r}^{2}-4{q}^{2}{l}^{2}+{q}^{4}+3{l}^{4} \right)  \Delta_r\,,
\]
\[
{\bar\omega}_4 = - 2{b}^{3}q  \left(4{l}^{2}m+{r}^{3}-3r{l}^{2} \right) \Delta_r\,,
\]
\[
{\bar\omega}_3 = -2 {b}^{2}l \left( {b}^{2}{l}^{4}-2{b}^{2}{l}^{2}{q}^{2}+6{b}^{2}{l}^{2}{
	r}^{2}-2{b}^{2}{l}^{2}mr-2{b}^{2}{q}^{2}{r}^{2}-3{q}^{2}+{b}^{2}{r}^{4}+2{b}^{2}{r}^{3}m \right)  \Delta_r\,,
\]
\[
{\bar\omega}_2 = 2bqr \left( {b}^{2}{q}^{2}{r}^{2}-3{b}^{2}{l}^{2}{q}^{2}-2{b}^{2
}{r}^{3}m-4{b}^{2}{l}^{2}{r}^{2}+{r}^{2}+6{b}^{2}{l}^{2}mr+4{b}^{2}{l}^{4}+{l}^{2} \right) \,,
\]
\[
{\bar\omega}_1 = -l \left( {b}^{4}{l}^{4}-6{l}^{2}{b}^{4}{r}^{2}-1-3{b}^{4}{r}^{4} \right)\Delta_r\,,
\]
\[
{\bar\omega}_0 = -2rqb \left( {r}^{2}+{l}^{2} \right)  \left( {b}^{2}{l}^{2}+1-{b}^{2}{r}^{2} \right) \,.
\]
Furthermore, the accompanying Maxwell vector field to this magnetized metric can be written in a compact form as
\be \label{Asol}
A'_\mu  {\rm{d}}x^\mu   = \frac{{{\rm{d}}t\sum\limits_{k = 0}^6 {\bar c_k x^k }  + {\rm{d}}\phi \sum\limits_{k = 0}^4 {\bar d_k x^k } }}{\sum\limits_{k = 0}^4 {\bar f_k x^k } }\,.
\ee 
The functions ${\bar c}_k$ and ${\bar d}_k$ appearing in the numerator of r.h.s. in eq. (\ref{Asol}) are 
\[
{\bar c}_6 = b^6 q \left\{{q}^{4}r- \left( 2{r}^{3}-8{l}^{2}m+10r{l}^{2} \right) {q}^{2}-12r{l}^{2}{m}^{2}+16{r}^{2}{l}^{2}m-2{r}^{3}{l}^{2}-8{l}^{4}m+{r}^{5}+9r{l}^{4}\right\} \Delta_r\,,
\]
\[
{\bar c}_5 = -2 b^5 l \left\{7{q}^{4}+ \left( 2{r}^{2}-12mr-10{l}^{2} \right) {q}^{2}+4{l
}^{2}{m}^{2}+4m{r}^{3}+6{r}^{2}{l}^{2}-4{l}^{2}rm-{r}^{4}+3{l}
^{4}\right\} \Delta_r\,,
\]
\[
{\bar c}_4 = q b^4 \left\{ \left( {b}^{2}r+8{b}^{2}m \right) {l}^{6}+ \left( 3r+5{b}^{2}{r
}^{3}+4{b}^{2}{r}^{2}m-8{q}^{2}{b}^{2}m+12{b}^{2}r{m}^{2}-10{q}^{2}{b}^{2}r-8m \right) {l}^{4} \right.
\]
\[
+ \left( \left( -5{r}^{5}+4{r}^{3}{m}^{2}-4{q}^{2}{r}^{3}+16m{r}^{4}-
20{q}^{2}{r}^{2}m+9{q}^{4}r \right) {b}^{2}+44m{r}^{2}-26{r}^{
	3}-28r{m}^{2}+6{q}^{2}r+8{q}^{2}m\right) {l}^{2}
\]
\[
\left. -{r}^{5}+3{q}^{4}{b}^{2}{r}^{3}+12{q}^{2}{r}^{2}m-4m{r}^{4}-2{q}^{2}{b}^{2}{r}^{5}-9{q}^{4}r+4{b}^{2}m{r}^{6}+2{q}^{2}{r}^{3}-{b}^{2}{r}^{7}-4{q}^{2}
{b}^{2}m{r}^{4} \right\}\,,
\]
\[
{\bar c}_3 = -4 b^3 l \left\{ \left(1-{b}^{2}{l}^{2}+5{b}^{2}{r}^{2} \right) {q}^{4}+ \left( 4r{b}^{2}{l}^{2}m-8mr+4{b}^{2}{r}^{4}-12{r}^{3}{b}^{2}m+6{r}^
{2}-12{b}^{2}{l}^{2}{r}^{2}-2{l}^{2} \right) {q}^{2} \right.
\]
\[
\left. +\left( 4{r}^{4}{m}^{2}-{r}^{6}-4{l}^{2}{r}^{2}{m}^{2}+16{l}^{2}{r}^{3}m-5{l}^{2}{r}^{4}+{l}^{6}+5{l}^{4}{r}^{2} \right) {b}^{2}+{r}^{4}-2{r}^{2}{l}^{2}+{l}^{4}+4{r}^{2}{m}^{2}+4{l}^{2}rm-4m{r}^{3} \right\}\,,
\]
\[
{\bar c}_2 = b^2 q r\left\{ \left( 3{l}^{2}{r}^{4}-20{l}^{2}{r}^{3}m-11{l}^{4}{r}^{2}+3{q}^{2}{r}^{4}+14{l}^{2}{r}^{2}{q}^{2}-9{l}^{4}{q}^{2}+6{l}^{4}rm+9{l}^{6}-{r}^{6}-2m{r}^{5} \right) {b}^{4} \right.
\]
\[
\left.  + \left(44{r}^{2}{l}^{2} -14{q}^{2}{r}^{2}+6{l}^{2}{q}^{2}-2{l}^{4}+6{r}^{4}+4m{r}^{3}-36{l}^{2}rm \right) {b}^{2}-9{q}^{2}-{r}^{2}+6mr+5{l}^{2} \right\}\,,
\]
\[
{\bar c}_1 = -2 b l\left\{ \left( 6{l}^{2}{r}^{2}{q}^{2}+3{q}^{2}{r}^{4}-{l}^{4}{q}^{2}+3{l}^{2}{r}^{4}-7{l}^{4}{r}^{2}+3{r}^{6}-6m{r}^{5}+{l}^{6}+2{l}^
{4}rm-12{l}^{2}{r}^{3}m \right) {b}^{4} \right.
\]
\[
\left.  + \left( 4{l}^{2}rm-4m{r}
^{3}-4{r}^{2}{l}^{2}-6{q}^{2}{r}^{2}-2{l}^{2}{q}^{2}+2{r}^{4}+
2{l}^{4} \right) {b}^{2}+2mr+{l}^{2}-{q}^{2}-{r}^{2} \right\}\,,
\]
\[
{\bar c}_0 = qr \left(1-5{b}^{6}{r}^{2}{l}^{4}-5{b}^{4}{r}^{4}+5{b}^{6}{r}^{4}{l}^{2}-18{b}^{4}{r}^{2}{l}^{2}-{l}^{6}{b}^{6}+{b}^{6}{r}^{6}+{b}^{2}{l}^{2}-5{b}^{2}{r}^{2}-{l}^{4}{b}^{4}\right)\,,
\]
and
\[
{\bar d}_4 = -b^3 \left\{ \left(12{q}^{2}{r}^{2} -32mr{q}^{2}-40m{r}^{3}+36{r}^{2}{m}^{2}+15{r}^{4}+9{q}^{4} \right) {l}^{2}+r^2\left(q^2-r^2\right)^2\right.
\]
\[
\left.  9{l}^{6}+ \left( 4{m}^{2}-18{q}^{2}+24mr-9{r}^{2} \right) {l}^{4} \right\}\,,
\]
\[
{\bar d}_3 = 6 b^2 lq \left\{3m{r}^{2}-{q}^{2}r+{l}^{2}m-2{r}^{3}\right\}\,,
\]
\[
{\bar d}_2 = b \left\{ \left( -12{l}^{2}{r}^{2}{q}^{2}+6{l}^{2}{r}^{4}+2{r}^{6}+8{l}
^{2}{r}^{3}m-2{q}^{2}{r}^{4}+6{l}^{4}{q}^{2}-6{l}^{6}-24{l}^{4
}rm+30{l}^{4}{r}^{2} \right) {b}^{2}\right.
\]
\[
\left.  +{r}^{4}+{l}^{2}{q}^{2}-8{l}^{2}rm+6{r}^{2}{l}^{2}-3{q}^{2}{r}^{2}-3{l}^{4} \right\}\,,
\]
\[
{\bar d}_1 = 2qrl \left( 1+6{b}^{2}{l}^{2} \right)\,,
\]
\[
{\bar d}_0 = -b \left( {l}^{2}+{r}^{2} \right)  \left( {b}^{2}{l}^{4}+{l}^{2}+6{b}^{2}{l}^{2}{r}^{2}+{b}^{2}{r}^{4}+{r}^{2} \right) \,,
\]
while the function  ${\bar f}_k $ in the denominator are
\[
{\bar f}_4 = b^4 \left\{4l^2m^2 \left(9r^2+l^2\right) + 8r{l}^{2} m \left( 3{l}^{2}-5{r}^{2}-4{q}^{2} \right) \right.
\]
\[
\left. + {r}^{6}+ \left( 15{l}^{2}-2{q}^{2} \right) {r}^{4}+ \left( {q}^{4}
+12{l}^{2}{q}^{2}-9{l}^{4} \right) {r}^{2}-18{l}^{4}{q}^{2}+9{l}^{6}+9{l}^{2}{q}^{4} \right\}\,,
\]
\[
{\bar f}_3= 8{b}^{3}lq \left\{{q}^{2}r-3m{r}^{2}-{l}^{2}m+2{r}^{3} \right\}\,,
\]
\[
{\bar f}_2 = 2b^2 \left\{ \left( 12{l}^{4}rm+3{l}^{6}-4{l}^{2}{r}^{3}m+{q}^{2}{r}^{4}+6{l}^{2}{r}^{2}{q}^{2}-3{l}^{2}{r}^{4}-15{l}^{4}{r}^{2}-3{l}^{4}{q}^{2}-{r}^{6} \right) {b}^{2} \right.
\]
\[
\left. -{r}^{4}+3{q}^{2}{r}^{2}+8{l}^{2}rm-6{r}^{2}{l}^{2}+3{l}^{4}-{l}^{2}{q}^{2}\right\}\,,
\]
\[
{\bar f}_1 = -8bqrl \left\{1+2{b}^{2}{l}^{2}\right\} \,,
\]
\[
{\bar f}_0 = \left( {l}^{2}+{r}^{2} \right)  \left\{ {l}^{4}{b}^{4}+6{b}^{4}{r}^{2}{l}^{2}+2{b}^{2}{l}^{2}+2{b}^{2}{r}^{2}+{b}^{4}{r}^{4}+1\right\}\,.
\]

The solutions (\ref{magLPWmetric}) and (\ref{Asol}), which would be referred as the magnetized \RN-Taub-NUT (MRNTN) solution, can be considered as the Taub-NUT extension of magnetized \RN black holes proposed in \cite{Ernst}, and studied further in \cite{Astorino:2015lca}. As we expect for a black hole immersed in magnetic field, the area of horizon is just the same to that of the non-magnetized one \cite{Siahaan:2015xia,Astorino:2015lca,Astorino:2015naa}. The similar situation is repeated here, where the radius of MRNTN black hole is just the same to that of RNTN spacetime, namely $r_\pm = m \pm \sqrt{m^2-q^2+l^2}$. Accordingly, the extremal conditions between the magnetized and non-magnetized RNTN black hole is also the same, i.e. $m^2+l^2 = q^2$. 

\section{Some aspects of the spacetime}\label{sec.aspects}

\subsection{Kretschmann scalar}

In studying a curved spacetime, Kretschmann scalar $K = {R_{\alpha \beta \mu \nu }}{R^{\alpha \beta \mu \nu }}$ can be used to identify the existence true singularity in the spacetime. It is well known that a spacetime with NUT parameter is regular at the origin, where the singularity lies on an axis instead. This type of singularity is known by a conic singularity, where the periodicity of $\phi$ coordinate is no longer $2\pi$. Note that the conic singularity in spacetime with NUT parameter differs to that of the magnetized spacetime, where the latter case can be cured by a scaling in $\phi$ coordinate \cite{Aliev:1989wx}. 

Obviously the full expression for Kretschmann scalar of MRNTN spacetime is lengthy, even on the equator $x=0$. Nevertheless, the regularity of this quantity at origin can be identified from its denominator, where this scalar takes the form
\be \label{K}
K \sim \frac{1}{\left({\sum\limits_{k = 0}^6 {{c_k}{l^k}} }\right)^6}\,,
\ee 
with
\[
c_6 = {b}^{4} \left( 3{x}^{2}+1 \right) ^{2}\,,
\]
\[
c_5 = 0\,,
\]
\[
c_4 = {b}^{2} \left( 4{b}^{2}{m}^{2}{x}^{4}+24{b}^{2}mr{x}^{4}-18{x}^{4}{q}^{2}{b}^{2}-9{b}^{2}{r}^{2}{x}^{4}+24{b}^{2}mr{x}^{2}\right.
\]
\[
\left. -6{b}^{2}{q}^{2}{x}^{2}-30{b}^{2}{r}^{2}{x}^{2}+7{b}^{2}{r}^{2}+6{x}^{2}+2 \right)\,,
\]
\[
c_3 =-8{b}^{3}qx \left( m{x}^{2}+2r \right)\,, 
\]
\[
c_2 =36{b}^{4}{m}^{2}{r}^{2}{x}^{4}-32{b}^{4}m{q}^{2}r{x}^{4}-40{b}^{4}m{r}^{3}{x}^{4}+9{b}^{4}{q}^{4}{x}^{4}+12{r}^{2}{b}^{4}{q}^{2}{x}^{4}+15{r}^{4}{b}^{4}{x}^{4}-8{b}^{4}m{r}^{3}{x}^{2}
\]
\[
+12{r}^{2}{b}^{4}{q}^{2}{x}^{2}-6{r}^{4}{b}^{4}{x}^{2}+7{r}^{4}{b}^{4}+16{b}^{2}mr{x}^{2}-2{b}^{2}{q}^{2}{x}^{2}-12{b}^{2}{r}^{2}{x}^{2}+4{b}^{2}{r}^{2}+1\,,
\]
\[
c_1 = -8bqrx \left( 3{b}^{2}mr{x}^{2}-{b}^{2}{q}^{2}{x}^{2}-2{b}^{2}{r}^{2}{x}^{2}+1 \right)\,,
\]
\[
c_0 ={r}^{2} \left( {b}^{4}{q}^{4}{x}^{4}-2{r}^{2}{b}^{4}{q}^{2}{x}^{4}+{r}^{4}{b}^{4}{x}^{4}+2{r}^{2}{b}^{4}{q}^{2}{x}^{2}-2{r}^{4}{b}^{4}
{x}^{2}+{r}^{4}{b}^{4}+6{b}^{2}{q}^{2}{x}^{2}-2{b}^{2}{r}^{2}{x}^{2}\right. 
\]
\[
\left. +2{b}^{2}{r}^{2}+1 \right)\,.
\]
Setting the external magnetic field parameter and electric charge vanish in eq. (\ref{K}), we are able to recover the Kretschmann scalar for Taub-NUT spacetime which reads
\[
K = \frac{48}{\left(r^2+l^2\right)^6}\left\{ \left( {l}^{4}-m{l}^{3}+3{l}^{3}r+3{l}^{2}mr-3{l}^{2}{r}^{2}+3m{r}^{2}l-{r}^{3}l-m{r}^{3} \right)  \right.
\]
\be 
\left. \times \left( {l}^{4}+m{l}^
{3}-3{l}^{3}r+3{l}^{2}mr-3{l}^{2}{r}^{2}-3m{r}^{2}l+{r}^{3}l-m
{r}^{3} \right) \right\}\,.
\ee  
In the last equation, the regularity at origin is obvious for the non-vanishing NUT parameter. 

\begin{figure}
	\centering
	\includegraphics[scale=0.6]{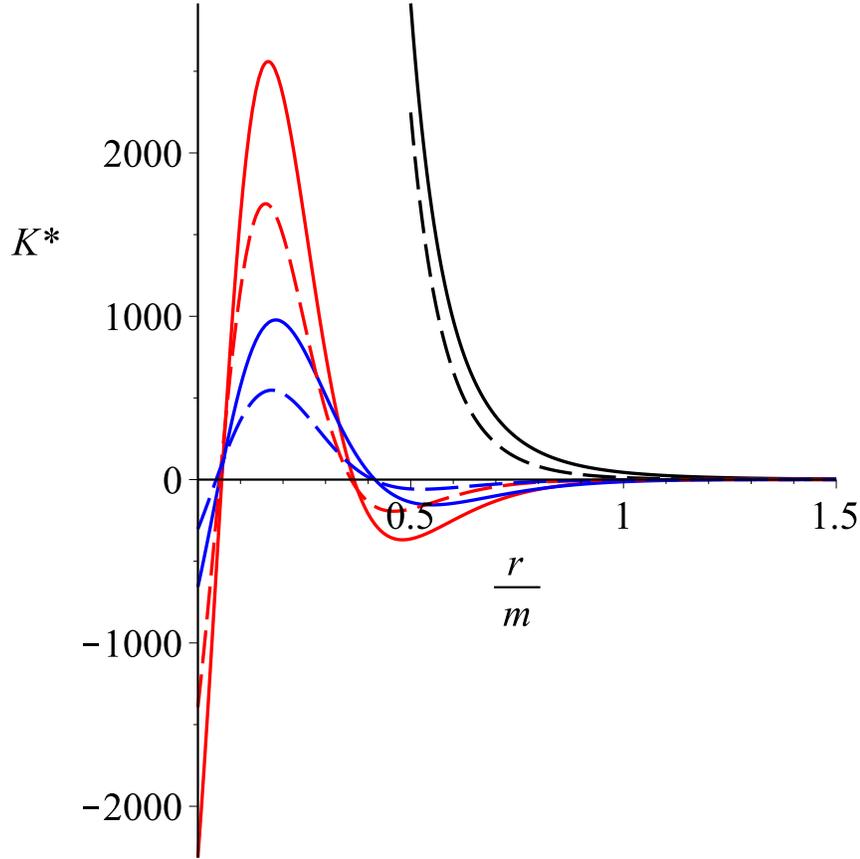}
	\caption{Plots of the dimensionless Kretschmann scalar for $q=0.1$, i.e. $K^* = m^4 R_{\alpha\beta\mu\nu} R^{\alpha\beta\mu\nu}$, with $bm = 0.1$ for solid lines and $bm = 0.5$ for dashed lines. The black lines are the cases of $l=0$, red lines are $l=0.5$, and blue lines are $l=0.6$.}\label{fig.plotK}
\end{figure}

To proceed with some numerical studies, let us consider the quantity (\ref{K}) on equator where the full expression is given in the appendix. Illustrations for this quantity evaluated for some numerical setups are given in fig. \ref{fig.plotK}. In the plots, we find that the typical behavior of spacetime with NUT parameter do appear, i.e. the corresponding Kretschmann scalar at origin is regular or finite. As comparison, we also provide the null case of NUT parameter, where the singularity at origin still exist in the spacetime with external magnetic field. 

\subsection{Electric charge and angular momentum}\label{subsec.QJ}

As a generalization of the \RN spacetime solution, there should be an electric charge associated to the MRNTN spacetime. Following \cite{Booth:2015nwa}, the electric charge can be found as
\be \label{Qtotal}
Q = \frac{{{{\left| {\Lambda \left( {x = 1} \right)} \right|}^2}}}{2}\left[ {{{\tilde A}_\phi }} \right]_{x = 1}^{x =  - 1} = \frac{{\sum\limits_{k = 0}^6 {{h_k}{r^k}} }}{{\sum\limits_{j = 0}^6 {{{\tilde h}_k}{r^k}} }} \,,
\ee 
where
\[
h_6 = -16\,{b}^{4}{l}^{2} q\left( 4\,{b}^{2}{l}^{2}+{b}^{2}{q}^{2}+3 \right) \,,
\]
\[
h_5 = 16\,q{b}^{4}{l}^{2}m \left( 16\,{b}^{2}{l}^{2}+3\,{b}^{2}{q}^{2}+9
\right)\,,
\]
\[
h_4 = q \left( 192{b}^{6}{l}^{6}-336{b}^{6}{l}^{4}{m}^{2}-112{b}^{6}{l}^{4}{q}^{2}-36{b}^{6}{l}^{2}{m}^{2}{q}^{2}-28{b}^{6}{l}^{2}{q}^{4}-{b}^{6}{q}^{6} +112{b}^{4}{l}^{4}\right. 
\]
\[
\left. -108{b}^{4}{l}^{2}{m}^{2}-88{b}^{4}{l}^{2}{q}^{2}-5{b}^{4}{q}^{4}+20{b}^{2}{l}^{2}+5{b}^{2}{q}
^{2}+1 \right) \,,
\]
\[
h_3 = -4q{b}^{2}{l}^{2}m \left( 128{b}^{4}{l}^{4}-36{b}^{4}{l}^{2}{m}^
{2}-80{b}^{4}{l}^{2}{q}^{2}-9{b}^{4}{q}^{4}+48{b}^{2}{l}^{2}-30
{b}^{2}{q}^{2}+7 \right)  \,,
\]
\[
h_2 = -4q{b}^{2}{l}^{2} \left( 48{b}^{4}{l}^{6}-80{b}^{4}{l}^{4}{m}^{2
}-68{b}^{4}{l}^{4}{q}^{2}+60{b}^{4}{l}^{2}{m}^{2}{q}^{2}+20{b}^{4}{l}^{2}{q}^{4} +3{b}^{4}{q}^{6}\right. 
\]
\[
\left. +20{b}^{2}{l}^{4}-8{b}^{2}{l}^{2}{m}^{2}-24{b}^{2}{l}^{2}{q}^{2}+10{b}^{2}{q}^{4}+4{l}^{2}-5{q}^{2} \right) \,,
\]
\[
h_1 = q{b}^{2}{l}^{4}m \left( 64{b}^{4}{l}^{4}+4{b}^{4}{l}^{2}{m}^{2}-92{b}^{4}{l}^{2}{q}^{2}+33{b}^{4}{q}^{4}+12{b}^{2}{l}^{2}-2{b}^{2}{q}^{2}+1 \right)  \,,
\]
\[
h_0 = q{l}^{4} \left( 64{b}^{6}{l}^{6}+16{b}^{6}{l}^{4}{m}^{2}-144{b}^
{6}{l}^{4}{q}^{2}-12{b}^{6}{l}^{2}{m}^{2}{q}^{2}+108{b}^{6}{l}^{2}
{q}^{4} -27{b}^{6}{q}^{6} \right. 
\]
\[
\left. +16{b}^{4}{l}^{4}+12{b}^{4}{l}^{2}{m}^{2}-8{b}^{4}{l}^{2}{q}^{2}-3{b}^{4}{q}^{4}-4{b}^{2}{l}^{2}-{b}^{2}
{q}^{2}-1 \right) \,,
\]
\[
{\tilde h}_6 = 16\,{b}^{4}{l}^{2}\,,
\]
\[
{\tilde h}_5 = -16\,{b}^{3}l \left( 3\,blm+q \right)\,,
\]
\[
{\tilde h}_4 = -16{b}^{4}{l}^{4}+36{b}^{4}{l}^{2}{m}^{2}+24{b}^{4}{l}^{2}{q}^{2}+{b}^{4}{q}^{4}+24{b}^{3}lmq-8{b}^{2}{l}^{2}+6{b}^{2}{q}^{2}+1\,,
\]
\[
{\tilde h}_3 = -8lb \left( 4{b}^{3}lm{q}^{2}+{b}^{2}{q}^{3}-2blm-q \right) \,,
\]
\[
{\tilde h}_2 = -2{l}^{2} \left( 8{b}^{4}{l}^{4}-20{b}^{4}{l}^{2}{m}^{2}-5{b}^{4}{q}^{4}-16{b}^{3}lmq-2{b}^{2}{q}^{2}-1 \right) \,,
\]
\[
{\tilde h}_1 = 8b{l}^{3} \left( 6{b}^{3}{l}^{3}m-4{b}^{3}lm{q}^{2}+2{b}^{2}{l}^{2}q-{b}^{2}{q}^{3}+2blm+q \right)\,,
\]
\[
{\tilde h}_0 = {l}^{4} \left( 16{b}^{4}{l}^{4}+4{b}^{4}{l}^{2}{m}^{2}-24{b}^{4}
{l}^{2}{q}^{2}+9{b}^{4}{q}^{4}+8{b}^{3}lmq+8{b}^{2}{l}^{2}-2{b}^{2}{q}^{2}+1 \right) \,.
\]
This expression is not what one would have expected from total charge calculation. Even after considering the two-surface $S_\infty$ at spacelike infinity, i.e. taking $r\to \infty$ for the result in (\ref{Qtotal}), we obtain $Q_{r\to \infty} = -q\left(3+b^2q^2+4b^2l^2\right)$ which is finite but again not as we would expect. However, taking $l\to 0$ from eq. (\ref{Qtotal}) gives $Q_{l\to 0}= q\left(1-b^2 q^2\right)$, exactly as the total charge of magnetized \RN case \cite{Booth:2015nwa}, without taking the $r\to \infty$ limit.  

Now let us turn the total angular momentum by using the Ernst potential \cite{Booth:2015nwa}, 
\be \label{Jtotal}
J = \frac{{{{\left| {\Lambda \left( {x = 1} \right)} \right|}^4}}}{8}\left[ {\Psi + 2 A_\phi {{\tilde A}_\phi }} \right]_{x = 1}^{x =  - 1} \,.
\ee 
It turns out the expression for $J$ above is tedious and we omit to present the full result here. Just like the outcome for $Q$ above, the obtained result for $J$ in (\ref{Jtotal}) is still a function of radius $r$. However, considering the two-surface $S_\infty$ for $J$ leads to a singular result. Nevertheless, setting the null NUT parameter in the last equation gives the expected result for the magnetized \RN spacetime, i.e. $J = -2bq^3 \left(1+b^2 q^2\right)$. 

\subsection{Hawking temperature}
Here we discuss some aspects related to the MRNTN spacetime presented in the previous section. These aspects are those which consider to have relations to what we will discuss in the next section, namely the extremal charge black hole/CFT correspondence inspired by the Kerr/CFT conjecture \cite{Guica:2008mu}. We start with the area of black hole by using the standard textbook formula,
\be \label{area}
{\cal A} = \int\limits_0^{2\pi } {d\phi \int\limits_{ - 1}^1 {dx\sqrt {g_{\phi \phi } g_{xx} } } }  = 4\pi \left(r_ + ^2 + l^2\right)\,. 
\ee
This is just the area of a generic RNTN black hole \cite{Pradhan:2013hqa}, and consequently the corresponding entropy reads \cite{Pradhan:2013hqa},
\be \label{entropy}
S = \frac{\cal A}{4} = \pi \left(r_ + ^2 + l^2\right) \,.
\ee 

Now let us turn to the Hawking temperature, which can be computed using several ways. However, obtaining the Hawking temperature $T_H = \tfrac{\kappa}{2\pi}$ using the surface gravity
\be 
\kappa  = \sqrt { - \frac{1}{2}\left( {\nabla _\mu  \xi _\nu  } \right)\left( {\nabla ^\mu  \xi ^\nu  } \right)} \,,
\ee 
is found to be troublesome, due to the complexity of spacetime metric. However the calculation turns out to be much simpler if we consider the complex path method in $(1+1)$ dimension \cite{Srinivasan:1998ty}, where the case of Taub-NUT black hole is worked out in \cite{Kerner:2006vu} and for a Vaidya black hole in \cite{Siahaan:2009qv}. Note that the calculation in \cite{Kerner:2006vu} applies to a general stationary and axial symmetric black hole spacetime metric
\be \label{metricKerner}
ds^2  =  - \tilde f\left( {r,x} \right)dt^2  + \frac{{dr^2 }}{{\tilde g\left( {r,x} \right)}} + \tilde C\left( {r,x} \right)h_{ij} \left( {r,x} \right)d\tilde x^i d\tilde x^j \,,
\ee
where $\tilde x^i  = \left[ {x,\tilde \phi } \right]$, and ${\tilde \phi} = \phi - \omega' t$, and the line element (\ref{magLPWmetric}) has this form. Now we consider the geodesic with a fixed $x=0$ and $d{\tilde \phi}=0$, which then yields only the $\left(t-r\right)$ sector in the metric that matters. In such consideration, the reading of massless Klein-Gordon equation $\nabla_\mu \nabla^\mu \Phi$ can be written as
\be \label{eqS}
\left( {\frac{{\partial S}}{{\partial r}}} \right)^2  = \frac{1}{{\tilde f\left( r \right)\tilde g\left( r \right)}}\left( {\frac{{\partial S}}{{\partial t}}} \right)^2 \,,
\ee 
after employing the Hamilton-Jacobi ansatz for the scalar field $\Phi = \exp\left[-iS\left(t,r\right)\right]$. Since the spacetime under discussion is stationary, it is allowed to consider
\be \label{Sansatz}
S\left( {t,r} \right) = Et + \tilde S\left( r \right)\,,
\ee 
to solve eq. (\ref{eqS}), which leads to the solution
\be 
S\left( {t,r} \right) = E\left( {t \pm \int\limits_0^r {\frac{{dr}}{{\sqrt {\tilde f\left( r \right)\tilde g\left( r \right)} }}} } \right)\,.
\ee 
Based on this solution, we have the ingoing and outgoing fields
\be \label{Phi.in}
\Phi _{{\rm{in}}}  = \exp \left[ { - iE\left( {t + \int\limits_0^r {\frac{{dr}}{{\sqrt {\tilde f\left( r \right)\tilde g\left( r \right)} }}} } \right)} \right] \,,
\ee 
and
\be \label{Phi.out}
\Phi _{{\rm{out}}}  = \exp \left[ { - iE\left( {t - \int\limits_0^r {\frac{{dr}}{{\sqrt {\tilde f\left( r \right)\tilde g\left( r \right)} }}} } \right)} \right]\,,
\ee 
respectively. By imposing that the probability of ingoing particle is unity, i.e. $
P_{{\mathop{\rm in}\nolimits} }  = \left| {\Phi _{{\rm{in}}} } \right|^2  = 1$, and by using the detailed balance principle 
\be 
{{P_{{\rm{out}}} }} = \exp \left(-{E}/{{T_H }}\right) {{P_{{\rm{in}}} }}\,,
\ee  
we finally can get the Hawking temperature
\be 
T_H  = \frac{1}{4}\left( {{\mathop{\rm Im}\nolimits} \int\limits_0^r {\frac{{dr}}{{\sqrt {\tilde f\left( r \right)\tilde g\left( r \right)} }}} } \right)^{ - 1} \,.
\ee 
Plugging the metric function (\ref{magLPWmetric}) into the last equation, we have
\be 
T_H  = \frac{1}{4}\left( {{\mathop{\rm Im}\nolimits} \int\limits_0^r {\frac{{\left( {r^2  + l^2 } \right)dr}}{{\left( {r - r_ +  } \right)\left( {r - r_ -  } \right)}}} } \right)^{ - 1} \,,
\ee 
which then gives us the Hawking temperature for a MRNTN black hole
\be \label{TH}
T_H  = \frac{{r_ +   - m  }}{{2\pi \left( {r_ + ^2  + l^2 } \right)}}\,.
\ee
Note that the temperature (\ref{TH}) is just the one for the generic RNTN black hole \cite{Pradhan:2013hqa}. This is as what we expected, that the Hawking temperature for a magnetized black hole is exactly the same to that of the non-magnetized one \cite{Siahaan:2015xia,Astorino:2015lca,Astorino:2015naa}.

\section{Microscopic entropy for the extermal MRNTN black hole}\label{sec.kerrcft}

In this section, we extend the magnetized \RN/CFT correspondence reported in \cite{Astorino:2015lca} to the case with the presence of NUT parameter. The first step is to obtain the near horizon geometry for a MRNTN black hole, which can be achieved by performing the transformation 
\be \label{nearhorizon.transf}
t \to \frac{{r_0 t}}{\lambda }~~,~~r \to r_e  + \lambda r_0 r~~,~~\phi  \to \phi  + \Omega _J^{ext} \frac{{r_0 }}{\lambda }t\,.
\ee 
In equation above $\Omega_J = \omega'\left(r_+\right)$, and $\Omega _J^{ext}$ is the corresponding quantity evaluated at extremality. 

Note that from eq. (\ref{eq.Psi}) there exist a gauge freedom for the twist potential, namely $\Psi'\to \Psi' +\Psi'_0$ for a constant $\Psi'_0$, which leaves the Ernst equations to be invariant. Recall that $\Psi'$ is the magnetized twist potential obeying
\be 
- i\nabla \Psi'  = \frac{{ f'^2 }}{\rho }\nabla \omega' + 2{\bar \Phi'} \nabla \Phi'
\ee 
where $f'$, $\omega'$ are the magnetized metric functions, and $\Phi'$ is the magnetized electromagnetic Ernst potential. We then apply the near horizon transformation (\ref{nearhorizon.transf}) above to the gauged MRNTN metric, with 
\be 
\Psi' _0  = \frac{{2ml\left( {1 + 2b^2 l^2 } \right)}}{{b^2 \left( {l^2  - m^2 } \right)}}\,.
\ee 
The resulting near horizon metric of an extremal MRNTN black hole is found to be
\be \label{nhmetric}
{\rm{d}}s^2  = \Gamma \left( x \right)\left\{  - r^2 {\rm{d}}t^2  + \frac{{{\rm{d}}r^2 }}{{r^2 }} + \alpha \left( x \right){{\rm{d}}x^2 } \right\} + \gamma \left( x \right)\left( {{\rm{d}}\phi  + kr{\rm{d}}t} \right)^2 \,,
\ee 
where $r_0 = \sqrt{m^2 + l^2}$,  $\alpha\left(x\right) =\Delta_x^{-1}$,
\be 
\Gamma \left( x \right) = \frac{{q^4 \left( {4b^2 \left( {l^2  - m^2 } \right)x^2  + q^2 \left( {1 + q^2 b^2 } \right)^2 } \right)}}{{\left( {l^2  - m^2 } \right)^2 }}\,,
\ee
\be
\gamma \left( x \right) = \frac{q^4 \Delta_x}{\Gamma\left(x\right)} \,,
\ee
and
\be\label{k} 
k = -\frac{{4q^3 b\left( {1 + q^2 b^2 } \right)}}{{m^2 - l^2  }}\,.
\ee
Moreover, the associated vector field is 
\be \label{nhA}
A_\mu  {\rm{d}}x^\mu   = L\left( x \right)\left( {{\rm{d}}\phi  + kr{\rm{d}}t} \right)\,,
\ee 
where
\be 
L\left( x \right) = \frac{{q^4 x\sqrt {q^2 \left( {1 + q^2 b^2 } \right)^2  - 4b^2 \left( {l^2  - m^2 } \right)^2 } }}{{\Gamma \left( x \right)}}\,.
\ee 

One can find that the near horizon geometry (\ref{nhmetric}) above possesses the $SL\left(2,{\mathbb R}\right)\times U\left(1\right)$ isometry. The $SL\left(2,{\mathbb R}\right)$ symmetry is generated by the Killing vectors
\be 
K _ -   = \partial _t \,,
\ee 
\be 
K _0  = t\partial _t  - r\partial _r \,,
\ee 
\be 
K _ +   = \left( {\frac{1}{{2r^2 }} + \frac{{t^2 }}{2}} \right)\partial _t  - tr\partial _r  - \frac{k}{r}\partial _\phi  \,,
\ee 
obeying $\left[ {K _0 ,K _ \pm  } \right] =  \pm K _ \pm   $ and $\left[ {K_ -  ,K_ +  } \right] = K_0 $, and the $U\left(1\right)$ symmetry is generated by $ \partial_\phi$. Interestingly, we find that the presence of NUT parameter $l$ does not break the $SL\left(2,{\mathbb R}\right)\times U\left(1\right)$ isometry of the near horizon geometry of a magnetized \RN black hole \cite{Astorino:2015lca}. Furthermore, this isometry is the hint that the Kerr/CFT prescription to recover the entropy of extremal black hole should apply also to the MRNTN case. 

On the other hand, the authors of \cite{Hartman:2008pb} managed to obtain the central charge for a class of near horizon spacetime and vector field solutions in Einstein-Maxwell theory by using the asymptotic symmetry group (ASG) method. This work was reviewed and then generalized to a class of gravitational theories in \cite{Compere:2012jk}. Interestingly, the near horizon metric (\ref{nhmetric}) and vector field (\ref{nhA}) fall into the category discussed in \cite{Hartman:2008pb}. Therefore, the ASG method to obtain the corresponding central charge derived in \cite{Hartman:2008pb} should apply to the case of MRNTN black hole studied in this paper. 

To ensure that the ASG method works for the extremal MRNTN black hole, first we need to consider the following boundary condition to the near horizon metric
\be 
h_{\mu \nu }  \sim \left( {\begin{array}{*{20}c}
		{{\cal O}\left( {r^2 } \right)} & {{\cal O}\left( {r^{ 1} } \right)} & {{\cal O}\left( {r^{ - 1} } \right)} & {\cal O}\left( r^{-2} \right)  \\
		{} & {{\cal O}\left(1\right)} & {{\cal O}\left( {r^{ - 1} } \right)} & {{\cal O}\left( {r^{ - 1} } \right)}  \\
		{} & {} & { {\cal O}\left( {r^{ - 1} } \right)} & {{\cal O}\left( {r^{ - 2} } \right)}  \\
		{} & {} & {} & {{\cal O}\left( r^{-3} \right)}  \\
\end{array}} \right)\,,
\ee 
and the condition below for the accompanying vector field
\be 
a_\mu  {\rm{d}}x^\mu   \sim {\cal O}\left( r \right){\rm{d}}t + {\cal O}\left( {r^{ - 1} } \right){\rm{d}}r + {\cal O}\left( 1 \right){\rm{d}}x + {\cal O} \left(r^{-2}\right) {\rm{d}}\phi \,.
\ee 
Accordingly, the most general diffeomorphisms preserving the boundary condition for the metric is
\be 
\zeta^\mu \partial_\mu   = \varepsilon \left( \phi  \right)\partial _\phi   - r\frac{{d\varepsilon \left( \phi  \right)}}{{d\phi }}\partial _r  + {\rm{subleading~term}}
\ee 
which may lead us the associated central charge \cite{Hartman:2008pb,Astorino:2015lca,Siahaan:2015xia}
\be \label{central.charge.gen}
c = c_{grav}  + c_{gauge} \,,
\ee
where
\be 
c_{grav}  = 3k\int\limits_{ - 1}^{ + 1} {dx\sqrt {\Gamma \left( x \right)\alpha \left( x \right)\gamma \left( x \right)} } \,,
\ee 
and
\be 
c_{gauge} = 0.
\ee 
Inserting the metric component (\ref{nhmetric}) into eq. (\ref{central.charge.gen}) gives
\be\label{central.charge}
c =  - \frac{{24q^5 b\left( {1 + q^2 b^2 } \right)}}{{q^2  - 2l^2 }}\,,
\ee 
and it agrees to the central charge associated the near horizon geometry of an extremal magnetized \RN black hole \cite{Astorino:2015lca} after taking $b=B/2$ and $l=0$. 

Before we can employ the Cardy formula in recovering the extremal black hole entropy, we need to get the associated near horizon temperature. Clearly the Hawking temperature (\ref{TH}) vanishes in extremal limit, which is typical for any other extremal black holes. However, since the Hawking temperature is measured by an observer at infinity, the near horizon temperature is not necessary to vanish even in extremal state. This temperature can be computed in the following way
\be  
T_\phi   = \mathop {\lim }\limits_{r_ +   \to m} \frac{{T_H }}{{\Omega _J^{{\rm{ext}}}  - \Omega _J }} =  - \left. {\frac{{{{\partial T_H } \mathord{\left/{\vphantom {{\partial T_H } {\partial r_ +  }}} \right.\kern-\nulldelimiterspace} {\partial r_ +  }}}}{{{{\partial \Omega _J } \mathord{\left/{\vphantom {{\partial \Omega _J } {\partial r_ +  }}} \right.\kern-\nulldelimiterspace} {\partial r_ +  }}}}} \right|_{r_ +   = m} \,,
\ee 
which gives us the Frolov-Thorne temperature near an extremal black hole under consideration.
For an extremal MRNTN black hole, this temperature is
\be\label{FTtemp}
T_\phi   = \frac{{2l^2  - q^2 }}{{8\pi q^3 b\left( {1 + b^2 q^2 } \right)}} = \frac{1}{{2\pi k}}\,,
\ee 
where the constant $k$ is given in (\ref{k}). This result is exactly what we look for so the Kerr/CFT correspondence can provide us the ``holographic'' entropy calculation of an extremal MRNTN black hole by using the Cardy formula,
\be
S_{\rm Cardy} = \frac{\pi^2}{3} c T_\phi \,.
\ee 
Plugging the central charge (\ref{central.charge}) and Frolov-Thorne temperature (\ref{FTtemp}) into the last equation gives us the entropy of an extremal MRNTN black hole,
\be 
S_{\rm ext.} = \frac{{\cal A}_{\rm ext.}}{4} = \pi q^2\,.
\ee
This entropy is equal to the extremal case of (\ref{entropy}), so we have recovered the entropy for the black hole by using the microscopic Cardy formula. 

\section{Conclusion}
\label{sec:conclusion}

In this paper, we have presented a new solution in Einstein-Maxwell theory describing the magnetized spacetime outside a charged mass equipped with NUT parameter. The magnetization procedure is performed by using the Ernst prescription, applied to the RNTN solution as the seed. As expected, some properties of the MRNTN black hole are just those of the non magnetized one, for example the area of horizon and the Hawking temperature. 

Inspired by the Kerr/CFT proposal for some magnetized black holes \cite{Siahaan:2015xia,Astorino:2015lca,Astorino:2015naa}, we extend the conjecture to the case of MRNTN case. To proceed, first we need to obtain the corresponding near horizon and accompanying vector solution in an extremal MRNTN geometry. It turns out that resulting near horizon metric and the vector field solution are compatible with the general form that is used in the asymptotic symmetry group method \cite{Hartman:2008pb}. Therefore, the general formula for the central charge and Frolov-Thorne temperature achieved in \cite{Hartman:2008pb} can apply. In section \ref{sec.kerrcft}, we recover the extremal entropy for a MRNTN black hole by using Cardy formula as prescribed by Kerr/CFT correspondence \cite{Guica:2008mu,Hartman:2008pb,Compere:2012jk}. 

The spacetime solution presented in this paper is a generalization to the novel solution reported in our previous work, namely the magnetized Taub-NUT spacetime \cite{Siahaan:2021ypk}. Obviously, the similar solution generating method should apply if one considers the Kerr-Taub-NUT solution as a seed. Discussing the extremal Kerr/CFT correspondence to the obtained magnetized Kerr-Taub-NUT black hole also worth our consideration, i.e. extending the works in \cite{Siahaan:2015xia,Astorino:2015naa} to the case with NUT parameter. Also we have showed a problem in getting the charges associated to the spacetime in section \ref{subsec.QJ}. It appears that the presence of NUT charge leads to an unexpected but finite result for the total electric charge in the spacetime. However, the angular momentum investigation gives us a singular value which a typical issue in the spacetime with NUT parameter \cite{Aliev:2008wv}. On the other hand, defining such conserved quantities in a magnetized spacetime is not a trivial problem as well \cite{Astorino:2016hls,Booth:2015nwa}. Finding the proper mass, angular momentum, and electric charged in a magnetized spacetime with NUT parameter is a project we would like to pursue in the future.

\section*{Acknowledgement}

I thank LPPM-UNPAR for supporting my works. I also thank Merry K. Nainggolan for her encouragement.

\appendix

\section{Near horizon geometry and the twist potential}\label{app.twist}

Note that one can ``gauge'' the twist potential $\Psi$ in (\ref{Ernst.grav}) by adding some constant $\Psi_0$ to it. It results some changes in the magnetized solutions, and in particular for the metric functions $f'\to f' +\Delta f'$ and $\omega' \to \omega' + \Delta\omega'$. These changes are 
\[
\Delta\omega ' = -\frac{2 b^3 \Psi_0 }{\Delta_x r^4 + 2l^2 \left(1-3x^2\right) r^2 + l^2 \left(l^2+3 l^2 x^2 - 4q^2 x^2\right)} \left\{2b\Delta_r \left( 2{l}^{2}m-3r{l}^{2}-{r}^{3} \right) x^3 \right.
\]
\be 
\left. +2lq \left( 2{q}^{2}+3{r}^{2}-4mr-{l}^{2} \right)  x^2 + b \left( 6r{l}^{2}-l \Psi_0 +2{r}^{3} \right)  x -2ql \left( {r}^{2}+{l}^{2} \right) \right\}\,,
\ee 
and
\be 
\Delta f' = -b^3 \Psi_0\frac{ \Xi_1 \Xi_2}{\Xi_3 \Xi_4 }\,,
\ee 
where
\[
\Xi_1 = \Delta_x {r}^{4}+ \left( 2{l}^{2}-6{l}^{2}{x}^{2} \right) {r}^{2}+8mr{l}^{2}{x}^{2}-4{q}^{2}{l}^{2}{x}^{2}+3{l}^{4}{x}^{2}+{l}^{4}\,,
\]
\[
\Xi_2 = 4bl \left( 3r{l}^{2}-{l}^{2}m-{r}^{3}-2{q}^{2}r+3m{r}^{2} \right)  x^2 + 4q \left(l^2 - r^2\right) x + b \left( {r}^{2}+{l}^{2} \right)  \left(\Psi_0 -4rl \right) \,,
\]
\[
\Xi_3 = \Delta_x^2 b^4 r^6 + {b}^{2} \left( 2{b}^{2}{x}^{2}{q}^{2}+7{l}^{2}{b}^{2}-6{b}^{2}{l}^{2}{x}^{2}+15{b}^{2}{l}^{2}{x}^{4}-2{b}^{2}{q}^{2}{x}^{4}-2{x}^{2}+2 \right) r^4
\]
\[
-4l{b}^{3} \left( 2ml{x}^{2}b+10bl{x}^{4}m+\Psi_0 {x}^{2}b+\Psi_0 b-4{x}^{3}q \right) r^3 + \left( 12{l}^{2}{b}^{4}{x}^{2}{q}^{2}+4{l}^{2}{b}^{2}\right.
\]
\[
+12{l}^{2}{b}^{4}{x}^{4}{q}^{2}-24l{b}^{3}m{x}^{3}q+{b}^{4}{x}^{4}{q}^{4}+12l{b}^{4}m{x}^{2}\Psi_0-12{b}^{2}{l}^{2}{x}^{2}+7{b}^{4}{l}^{4} 
\]
\[
\left. +{\Psi_0}^{2}{b}^{4}+6{b}^{2}{x}^{2}{q}^{2}-9{l}^{4}{b}^{4}{x}^{4}-4{b}^{3}qxC+1-30{l}^{4}{b}^{4}{x}^{2}+36{l}^{2}{b}^{4}{m}^{2}{x}^{4}\right) r^2
\]
\[
+4bl \left( 2l{b}^{3}m \left\{3{l}^{2} -4{q}^{2}\right\}  {x}^{4}+2{b}^{2}{x}^{3}{q}^{3}+ \left\{4lmb+3{b}^{3} {\Psi_0} {l}^{2}+6{l}^{3}{b}^{3}m-2{b}^{3}{q}^{2} {\Psi_0}\right\}{x}^{2}\right.
\]
\[
\left. - \left\{2q+4{l}^{2}q{b}^{2}\right\} x-{b}^{3} {\Psi_0} {l}^{2}\right) r + {l}^{2}{b}^{4}{{\Psi_0}}^{2}+ \left( 4{l}^{2}{b}
^{3}qx-4{l}^{3}{b}^{4}m{x}^{2} \right) {\Psi_0} +{l}^{2}
\]
\[
-18{l}^{4}{b}^{4}{x}^{4}{q}^{2}-2{b}^{2}{l}^{2}{x}^{2}{q}^{2}+6{l}^{6}{b}^{4}{x}^{2} +2{b}^{2}{l}^{4}+6{b}^{2}{l}^{4}{x}^{2}+9{l}^{2}{b}^{4}{x}^{4}{q}^{4}
\]
\[
+4{l}^{4}{b}^{4}{m}^{2}{x}^{4}-8{l}^{3}{b}^{3}q{x}^{3}m+{l}^{6}{b}^{4}+9{l}^{6}{b}^{4}{x}^{4}-6{l}^{4}{b}^{4}{x}^{2}{q}^{2}\,,
\]
\[
\Xi_4 = \Xi_3 - b^3 \Psi_0 \left\{4bl \left( 3r{l}^{2}-{l}^{2}m-{r}^{3}-2{q}^{2}r+3m{r}^{2} \right) x^2 + 4q\left(r^2 - l^2\right) x \right.
\]
\[
\left. +b\left(r^2+l^2\right) \left(\Psi_0 - 4rl\right)\right\}\,.
\]

\section{Kretschmann scalar for MRNTN spacetime at $x=0$}

The Kretschmann scalar for MRNTN spacetime evaluated at equator, i.e. $x=0$, can be expressed as
\[
K_{x=0}  =\frac{\sum\limits_{j = 0}^{12} {{k_{2j}}{l^{2j}}}}{\left( {l}^{2}+{r}^{2} \right) ^{6} \left({b}^{4}{l}^{4}+6{b}^{4}{l}^{2}{r}^{2}+{b}^{4}{r}^{4}+2{b}^{2}{l}^{2}+2{b}^{2}{r}^{2}+1\right) ^{6}\,,
}
\]
where
\[
k_0 = 8{r}^{4} \left( {b}^{2}{r}^{2}+1 \right) ^{4} \left( 126{b}^{8}{m}
^{2}{r}^{10}-144{b}^{8}m{q}^{2}{r}^{9}-108{b}^{8}m{r}^{11} +43{r}
^{8}{q}^{4}{b}^{8}+60{r}^{10}{q}^{2}{b}^{8}+24{b}^{8}{r}^{12}+86{r}^{4}{q}^{4}{b}^{4} \right.
\]
\[
-216
{b}^{6}{m}^{2}{r}^{8}+676{b}^{6}m{q}^{2}{r}^{7}+204{b}^{6}m{r}^{
	9}-264{r}^{6}{q}^{4}{b}^{6}-356{r}^{8}{q}^{2}{b}^{6}-48{b}^{6}{r
}^{10}+124{b}^{4}{m}^{2}{r}^{6}-36{b}^{4}m{q}^{2}{r}^{5} -148{b}^
{4}m{r}^{7}
\]
\[
\left. +36{b}^{4}{q}^{2}{r}^{6}+40{b
}^{4}{r}^{8}+24{b}^{2}{m}^{2}{r}^{4}+28{b}^{2}m{q}^{2}{r}^{3}-12
{b}^{2}m{r}^{5}-48{r}^{2}{q}^{4}{b}^{2}+4{b}^{2}{q}^{2}{r}^{4}+6
{m}^{2}{r}^{2}-12m{q}^{2}r+7{q}^{4} \right) \,,
\]
\[
k_2 = -16{r}^{2} \left( {b}^{2}{r}^{2}+1 \right) ^{2} \left( 261{b}^{12}
{m}^{2}{r}^{14}-204{b}^{12}m{q}^{2}{r}^{13}-390{b}^{12}m{r}^{15}+
17{b}^{12}{q}^{4}{r}^{12}+186{b}^{12}{q}^{2}{r}^{14} \right.
\]
\[
-1230{b}^{10}{m}^{2}{r}^{12}-2034{b}^{10}m{q}^{2}{r}^{11}+
1152{b}^{10}m{r}^{13}+1148{b}^{10}{q}^{4}{r}^{10}+1166{b}^{10}{q
}^{2}{r}^{12}-276{b}^{10}{r}^{14}+1963{b}^{8}{m}^{2}{r}^{10}
\]
\[
-7844
{b}^{8}m{q}^{2}{r}^{9}-1608{b}^{8}m{r}^{11}+2367{r}^{8}{q}^{4}{b
}^{8}+3770{r}^{10}{q}^{2}{b}^{8}+353{b}^{8}{r}^{12}-972{b}^{6}{m
}^{2}{r}^{8}+3124{b}^{6}m{q}^{2}{r}^{7}+1184{b}^{6}m{r}^{9}
\]
\[
-1792
{r}^{6}{q}^{4}{b}^{6}-1972{r}^{8}{q}^{2}{b}^{6}-304{b}^{6}{r}^{10}
+67{b}^{4}{m}^{2}{r}^{6}+780{b}^{4}m{q}^{2}{r}^{5}-174{b}^{4}m{r}^{7}+1007{r}^{4}{q}^{4}{b}^{4}-482{b}^{4}{q}^{2}{r}^{6}+71{b}^{4}{r}^{8}
\]
\[
 +286{b}^{2}m{q}^{2}{r}^{3}-192
{b}^{2}m{r}^{5}-300{r}^{2}{q}^{4}{b}^{2}-42{b}^{2}{q}^{2}{r}^{4}+
36{b}^{2}{r}^{6}+45{m}^{2}{r}^{2}-60m{q}^{2}r-36m{r}^{3}
\]
\[
\left. +117{b}^{12}
{r}^{16}+186{b}^{2}{m}^{2}{r}^{4}+17{q}^{4}+30{q}^{2}{r}^{2}+3{r}^{4} \right) \,,
\]
\[
k_4 = \left( 76992m{q}^{2}r -74544{m}^{2}{r}^{2}+68832m{r}^{3}-18056{q}^{4}-35616{q}^{2}{r}^{2}-15120{r}^{4} \right) {r^{16}b^{16}}
\]
\[
+\left( 43584{m}^{2}{r}^{2}+304128m{q}^{2}r-24096m{r}^{3}-139008
{q}^{4}-162048{q}^{2}{r}^{2}+1536{r}^{4} \right) {r^{14}b^{14}}
\]
\[
+ \left( 25216{m}^{2}{r}^{2}+860736m{q}^{2}r-43136m{r}^{3}-292864
{q}^{4}-401216{q}^{2}{r}^{2}+7520{r}^{4} \right) {r^{12}b^{12}}
\]
\[
+\left(142240m{q}^{2}r-156992{m}^{2}{r}^{2}+58752m{r}^{3}+
52896{q}^{4}+23776{q}^{2}{r}^{2}-8384{r}^{4} \right) {r^{10}b^{10}}
\]
\[
+ \left(11840{r}^{4} -13472{m}^{2}{r}^{2}-821024m{q}^{2}r-46752m{r}^{3}+
134992{q}^{4}+434464{q}^{2}{r}^{2} \right) {r^{8}b^{8}}
\]
\[
+ \left(12512m{r}^{3} -28480{m}^{2}{r}^{2}-230080m{q}^{2}r-67008
{q}^{4}+67264{q}^{2}{r}^{2}-2944{r}^{4} \right) {r^{6}b^{6}}
\]
\[
+ \left( 3232{r}^{4}
-23424{m}^{2}{r}^{2}+2048m{q}^{2}r+6080m{r}^{3}+72960{q}^{4}-
23168{q}^{2}{r}^{2} \right) {r^{4}b^{4}}
\]
\[
+ \left( 3648{r}^{4} -960{m}^
{2}{r}^{2}+16672m{q}^{2}r-6720m{r}^{3}-7136{q}^{4}-10912{q}^{2
}{r}^{2}\right) {r^{2}b^{2}}
\]
\[
+720{m}^{2}{r}^{2}-480m{q}^{2}r-1920m{r}^{3}+56{q}^{4}+960{q}^{2}{r}^{2}+720{r}^{4}\,,
\]
\[
k_6 = -282096{b}^{16}{m}^{2}{r}^{16}+302592{b}^{16}m{q}^{2}{r}^{15}+
184608{b}^{16}m{r}^{17}-75968{b}^{16}{q}^{4}{r}^{14}-87840{b}^{16}{q}^{2}{r}^{16}-30576{b}^{16}{r}^{18}
\]
\[
+328704{b}^{14}{m}^{2}{r}^
{14}+788224{b}^{14}m{q}^{2}{r}^{13}-240384{b}^{14}m{r}^{15}-361472
{b}^{14}{q}^{4}{r}^{12}-386816{b}^{14}{q}^{2}{r}^{14}+26592{b}^{14}{r}^{16}
\]
\[
+111744{b}^{12}{m}^{2}{r}^{12}+550464{b}^{12}m{q}^{2}{r}^{11}-159104{b}^{12}m{r}^{13}+48768{b}^{12}{q}^{4}{r}^{10}-116544
{b}^{12}{q}^{2}{r}^{12}+3072{b}^{12}{r}^{14}
\]
\[
-247488{b}^{10}{m}^{2}{r}^{10}-2836512{b}^{10}m{q}^{2}{r}^{9}-102656{b}^{10}m{r}^{11}+859872{b}^{10}{q}^{4}{r}^{8}+1431968{b}^{10}{q}^{2}{r}^{10}+24896
{b}^{10}{r}^{12}
\]
\[
+106656{b}^{8}{m}^{2}{r}^{8}-1326080{b}^{8}m{q}^
{2}{r}^{7}-253472{b}^{8}m{r}^{9}-185408{b}^{8}{q}^{4}{r}^{6}+
241920{b}^{8}{q}^{2}{r}^{8}+40448{b}^{8}{r}^{10}
\]
\[
+26240{b}^{6}{m}
^{2}{r}^{6}+133440{b}^{6}m{q}^{2}{r}^{5}-104832{b}^{6}m{r}^{7}+
433984{b}^{6}{q}^{4}{r}^{4}-209984{b}^{6}{q}^{2}{r}^{6}+25952{b}
^{6}{r}^{8}
\]
\[
+15104{b}^{4}{m}^{2}{r}^{4}+78528{b}^{4}m{q}^{2}{r}^{3}
-54464{b}^{4}m{r}^{5}-58496{b}^{4}{q}^{4}{r}^{2}-49472{b}^{4}{q}
^{2}{r}^{4}+15424{b}^{4}{r}^{6}
\]
\[
+3648{b}^{2}{m}^{2}{r}^{2}-5408{b
}^{2}m{q}^{2}r-8064{b}^{2}m{r}^{3}+736{b}^{2}{q}^{4}+12960{b}^{2
}{q}^{2}{r}^{2}+384{b}^{2}{r}^{4}-48{m}^{2}+576mr-96{q}^{2}-
720{r}^{2}\,,
\]
\[
k_8 = 48+ \left(268032m{q}^{2}{r}^{13} -277920{m}^{2}{r}^{14}-14784m{
	r}^{15}-57040{q}^{4}{r}^{12}+81600{q}^{2}{r}^{14}+10512{r}^{16}
\right) {b}^{16}
\]
\[
+ \left( 1483776{m}^{2}{r}^{12}-400064m{q}^{2}{r}
^{11}-862272m{r}^{13}+167488{q}^{4}{r}^{10}+39360{q}^{2}{r}^{12}
+32256{r}^{14} \right) {b}^{14}
\]
\[
+ \left( 431488{m}^{2}{r}^{10}-
4368128m{q}^{2}{r}^{9}-336448m{r}^{11}+1931392{q}^{4}{r}^{8}+
1885312{q}^{2}{r}^{10}-73600{r}^{12} \right) {b}^{12}
\]
\[
+ \left( 
353472{m}^{2}{r}^{8}-3844768m{q}^{2}{r}^{7}-778176m{r}^{9}+
147296{q}^{4}{r}^{6}+474656{q}^{2}{r}^{8}+66112{r}^{10} \right) 
{b}^{10}
\]
\[
+ \left( 365344{m}^{2}{r}^{6}+363296m{q}^{2}{r}^{5}-356704
m{r}^{7}+1202000{q}^{4}{r}^{4}-846368{q}^{2}{r}^{6}-38080{r}^{8} \right) {b}^{8}
\]
\[
+ \left(51328{m}^{2}{r}^{4}+158304m{q}^{2}{r}^{
	3}-50336m{r}^{5}-224416{q}^{4}{r}^{2}-25056{q}^{2}{r}^{4}-35200
{r}^{6} \right) {b}^{6}
\]
\[
+ \left( 3968{m}^{2}{r}^{2}-17952m{q}^{2}
r+4704m{r}^{3}+3584{q}^{4}+51936{q}^{2}{r}^{2}-19712{r}^{4}
\right) {b}^{4}
\]
\[
+ \left( 2976mr-192{m}^{2}-1024{q}^{2}-3648{r
}^{2} \right) {b}^{2}\,,
\]
\[
k_{10} = 32{b}^{2} \left( 25599{b}^{14}{m}^{2}{r}^{12}-34644{b}^{14}m{q}^
{2}{r}^{11}-24930{b}^{14}m{r}^{13}+11277{b}^{14}{q}^{4}{r}^{10}+
20238{b}^{14}{q}^{2}{r}^{12} \right.
\]
\[
+1071{b}^{14}{r}^{14}+88212{b}^{12}{m}^{2}{r}^{10}-138726{b}^{12}m{q}^{2}{r}^{9}-12444{b}^{12}m{r}^{11}+61434{b}^{12}{q}^{4}{r}^{8}+25294{b}^{12}{q}^{2}{r}^{10}
\]
\[
-11670
{b}^{12}{r}^{12}+21840{b}^{10}{m}^{2}{r}^{8}-163042{b}^{10}m{q}^{2}{r}^{7}+2714{b}^{10}m{r}^{9}+36932{b}^{10}{q}^{4}{r}^{6}-18{b}^
{10}{q}^{2}{r}^{8}-10966{b}^{10}{r}^{10}
\]
\[
+15494{b}^{8}{m}^{2}{r}^{6}-7751{b}^{8}m{q}^{2}{r}^{5}-2072{b}^{8}m{r}^{7}+54377{b}^{8}{q}
^{4}{r}^{4}-52197{b}^{8}{q}^{2}{r}^{6}-11380{b}^{8}{r}^{8}+2327{b}^{6}{m}^{2}{r}^{4}
\]
\[
+7240{b}^{6}m{q}^{2}{r}^{3}+10161{b}^{6}m{r}^{5}-14855{b}^{6}{q}^{4}{r}^{2}+3912{b}^{6}{q}^{2}{r}^{4}-8180{b}^
{6}{r}^{6}-314{b}^{4}{m}^{2}{r}^{2}-693{b}^{4}m{q}^{2}r
\]
\[
\left. +2254{b}^
{4}m{r}^{3}+287{b}^{4}{q}^{4}+3057{b}^{4}{q}^{2}{r}^{2}-1667{b}^
{4}{r}^{4}-6{b}^{2}{m}^{2}+157{b}^{2}mr-114{b}^{2}{q}^{2}-121{b}^{2}{r}^{2}+9 \right) \,,
\]
\[
k_{12} = 16{b}^{4} \left( 131850{b}^{12}{m}^{2}{r}^{10}-175368{b}^{12}m{q
}^{2}{r}^{9}-19044{b}^{12}m{r}^{11}+56531{b}^{12}{q}^{4}{r}^{8}+
26364{b}^{12}{q}^{2}{r}^{10} \right.
\]
\[
-23898{b}^{12}{r}^{12}+91464{b}^{10}
{m}^{2}{r}^{8}-223280{b}^{10}m{q}^{2}{r}^{7}+174924{b}^{10}m{r}^{9}+91072{b}^{10}{q}^{4}{r}^{6}-102512{b}^{10}{q}^{2}{r}^{8}
\]
\[
-61008
{b}^{10}{r}^{10}-13640{b}^{8}{m}^{2}{r}^{6}-94884{b}^{8}m{q}^{2}{r}^{5}+95672{b}^{8}m{r}^{7}+83840{b}^{8}{q}^{4}{r}^{4}-108460{b}^
{8}{q}^{2}{r}^{6}-33652{b}^{8}{r}^{8}
\]
\[
+1156{b}^{6}{m}^{2}{r}^{4}+
27486{b}^{6}m{q}^{2}{r}^{3}+42360{b}^{6}m{r}^{5}-37042{b}^{6}{q}
^{4}{r}^{2}-9734{b}^{6}{q}^{2}{r}^{4}-18916{b}^{6}{r}^{6}-1566{b}^{4}{m}^{2}{r}^{2}
\]
\[\left.
-262{b}^{4}m{q}^{2}r+9970{b}^{4}m{r}^{3}+875{b}^{4}{q}^{4}+9414{b}^{4}{q}^{2}{r}^{2}+92{b}^{4}{r}^{4}+12{b}^{2}{m}^{2}-94{b}^{2}mr-432{b}^{2}{q}^{2}+792{b}^{2}{r}^{2}+50\right) \,,
\]
\[
k_{14}= 32{b}^{6} \left( 29529{b}^{10}{m}^{2}{r}^{8}-46272{b}^{10}m{q}^{2}{r}^{7}+74898{b}^{10}m{r}^{9}+17114{b}^{10}{q}^{4}{r}^{6}-41874
{b}^{10}{q}^{2}{r}^{8}-23103{b}^{10}{r}^{10} \right.
\]
\[
-13584{b}^{8}{m}^{2}
{r}^{6}-36336{b}^{8}m{q}^{2}{r}^{5}+91176{b}^{8}m{r}^{7}+15912{b
}^{8}{q}^{4}{r}^{4}-50672{b}^{8}{q}^{2}{r}^{6}+3894{b}^{8}{r}^{8}-
4964{b}^{6}{m}^{2}{r}^{4}
\]
\[
+19974{b}^{6}m{q}^{2}{r}^{3}+8308{b}^{6
}m{r}^{5}-13588{b}^{6}{q}^{4}{r}^{2}-26766{b}^{6}{q}^{2}{r}^{4}+
11496{b}^{6}{r}^{6}-242{b}^{4}{m}^{2}{r}^{2}+261{b}^{4}m{q}^{2}r
+3304{b}^{4}m{r}^{3}
\]
\[
\left. +413{b}^{4}{q}^{4}+8511{b}^{4}{q}^{2}{r}^{2}
+4550{b}^{4}{r}^{4}+15{b}^{2}{m}^{2}-411{b}^{2}mr-280{b}^{2}{q}^{2}+1448{b}^{2}{r}^{2}+25 \right) \,,
\]
\[
k_{16} = 8{b}^{8} \left( 534{b}^{8}{m}^{2}{r}^{6}-11280{b}^{8}m{q}^{2}{r}
^{5}+195300{b}^{8}m{r}^{7}+8879{b}^{8}{q}^{4}{r}^{4}-140052{b}^{8}{q}^{2}{r}^{6}+56124{b}^{8}{r}^{8}\right.
\]
\[
-30048{b}^{6}{m}^{2}{r}^{4} +57044{b}^{6}m{q}^{2}{r}^{3}+5484{b}^{6}m{r}^{5}-21772{b}^{6}{q}^
{4}{r}^{2}-98596{b}^{6}{q}^{2}{r}^{4}+126480{b}^{6}{r}^{6}+2944{b}^{4}{m}^{2}{r}^{2}-300{b}^{4}m{q}^{2}r
\]
\[
\left. -21036{b}^{4}m{r}^{3}+952
{b}^{4}{q}^{4}+44724{b}^{4}{q}^{2}{r}^{2}+26392{b}^{4}{r}^{4}+24
{b}^{2}{m}^{2}-1124{b}^{2}mr-1152{b}^{2}{q}^{2}+6072{b}^{2}{r}
^{2}-48 \right) \,,
\]
\[
k_{18} = -16{b}^{10} \left( 6345{b}^{6}{m}^{2}{r}^{4}-7548{b}^{6}m{q}^{2}
{r}^{3}-10782{b}^{6}m{r}^{5}+1841{b}^{6}{q}^{4}{r}^{2}+12450{b}^{6}{q}^{2}{r}^{4}-35847{b}^{6}{r}^{6}\right.
\]
\[
-1524{b}^{4}{m}^{2}{r}^{2}+
590{b}^{4}m{q}^{2}r+19188{b}^{4}m{r}^{3}-154{b}^{4}{q}^{4}-15030
{b}^{4}{q}^{2}{r}^{2}-13794{b}^{4}{r}^{4}
\]
\[
\left. +12{b}^{2}{m}^{2}-494
{b}^{2}mr+444{b}^{2}{q}^{2}+1334{b}^{2}{r}^{2}+58 \right)\,,
\]
\[
k_{20} = 8{b}^{12} \left( 882{b}^{4}{m}^{2}{r}^{2}-456{b}^{4}m{q}^{2}r-
16404{b}^{4}m{r}^{3}+43{b}^{4}{q}^{4}+7788{b}^{4}{q}^{2}{r}^{2}\right.
\]
\[
\left. +
16758{b}^{4}{r}^{4}-24{b}^{2}{m}^{2}+1452{b}^{2}mr-416{b}^{2}{q}^{2}-8736{b}^{2}{r}^{2}+28 \right) 
\]
\[
k_{22} = -48{b}^{14} \left( {b}^{2}{m}^{2}-78{b}^{2}mr+14{b}^{2}{q}^{2}+
681{b}^{2}{r}^{2}-18 \right) \,,
\]
\[
k_{24} = 336 {b}^{16}\,.
\]

\end{document}